\documentclass[prd,noshowpacs,nofootinbib,twocolumn]{revtex4-2}

\usepackage{color}
\usepackage{amssymb}
\usepackage{amsmath}
\usepackage{graphicx}
\usepackage{xcolor}
\usepackage{dcolumn}
\usepackage{amssymb}

\usepackage{amsfonts} 
\usepackage{psfrag}
\usepackage{appendix}
\usepackage{hyperref} 
\usepackage{mciteplus}
\usepackage{subcaption} 
\usepackage{caption}
\usepackage{bbold}
\usepackage{slashed}
\usepackage{mathbbol} 
\usepackage{amsthm}
\usepackage{mathrsfs}
\usepackage{setspace}
\usepackage{orcidlink}
\usepackage{enumitem}

\usepackage[T1]{fontenc}

\hypersetup{
colorlinks = true,
linkcolor = blue,
citecolor= cyan 
}

\usepackage[utf8]{inputenc}  
\usepackage{CJKutf8}    

\usepackage{pgfplots}
\pgfplotsset{compat=1.17}

\DeclareMathAlphabet{\mathpzc}{OT1}{pzc}{m}{it}
\usepackage{calligra}
\DeclareMathAlphabet{\mathcalligra}{T1}{calligra}{m}{n}
\DeclareFontShape{T1}{calligra}{m}{n}{<->s*[2.2]callig15}{}

\def\be {\begin{equation}}
\def\ee {\end{equation}}
\def\bea {\begin{eqnarray}}
\def\eea {\end{eqnarray}}
\def\bc {\begin{center}}
\def\ec {\end{center}}
\def\bfg {\begin{figure}}
\def\efg {\end{figure}}
\def\bi {\begin{itemize}}
\def\ei {\end{itemize}}

\DeclareMathOperator{\Area}{Area}

\def\beq{\begin{equation}}
\def\eeq{\end{equation}}
\def\br{\begin{eqnarray}}
\def\er{\end{eqnarray}}
\newcommand{\eel}[1] {\label{#1}\end{equation}}

\newcommand{\bdm}{\begin{displaymath}}
\newcommand{\edm}{\end{displaymath}}

\begin{document}

\title{Deriving the Cosmological Constant and Nature’s Constants from SU(3) Confinement Volume}

\author{Malak~Ali$^\S$}
\email{malak.ali.1113@brssd.org}

\author{Ahmed Farag Ali\orcidlink{0000-000X-XXXX-XXXX}$^{\P,\Vert}$}
\email{aali29@essex.edu}

\affiliation{$^\S$Ralston Middle School – 2675 Ralston Ave, Belmont, CA 94002, United States}
\affiliation{$^\P$Essex County College – 303 University Ave, Newark, NJ 07102, United States}
\affiliation{$^\Vert$Department of Physics, Benha University – Benha 13518, Egypt}

\date{\today}

\begin{abstract}
We explore the interplay between three well-established physical principles: QCD confinement, the Third Law of Thermodynamics, and holography, and examine how their combined implications may shed light on several open problems in fundamental physics. As the Universe cools toward absolute zero, color confinement fragments the vacuum into proton-scale domains. The Third Law, which renders $T = 0$ unattainable and prohibits complete entropy elimination, implies that these domains must persist. This leads to a natural count $N \simeq (R_u / R_p)^3 \sim 10^{123}$, where $R_u$ is the radius of the cosmic horizon and $R_p$ is the proton radius. Assuming holographic saturation, each domain corresponds to a Planck-area patch on the cosmic horizon, allowing the Planck length, and hence $\hbar$, $G$, and $c$, to emerge as geometric quantities. The same tiling dilutes the bare Planck-scale vacuum energy by a factor of $N$, reproducing the observed value of $\rho_\Lambda$ without requiring fine-tuned counterterms. 
\end{abstract}

\maketitle
\section{Introduction}\label{sec:intro}

The cosmological constant problem (CCP) represents one of the deepest unresolved tensions between quantum field theory, general relativity, and cosmology. Quantum field theory predicts a vacuum energy density of the order $\rho_{\text{QFT}} \sim M_{\text{Pl}}^4 \approx 10^{76}\,\text{GeV}^4$ \cite{Weinberg1989, Peebles2003}, whereas observational evidence from supernovae, baryon acoustic oscillations, and the cosmic microwave background consistently points to a vacuum energy density of $\rho_{\Lambda} \sim 10^{-47}\,\text{GeV}^4$ \cite{Riess1998, Perlmutter1999, Planck2018}.
This CCP has remained the
sharpest quantitative clash between gravity and quantum physics since
it was codified by Zel’dovich \cite{Zeldovich:1968ehl} and later reviewed by
Weinberg~\cite{Weinberg1989}.
A wide range of ideas have been pursued:  
cancellations by new symmetries or supersymmetry;  
infrared sequestering and screening mechanisms
(e.g.\ Kaloper–Padilla);  
dynamical relaxation through rolling scalar fields or
vacuum energy sequester
(see~\cite{Burgess2013,Padmanabhan2005} for reviews);  
and anthropic/environmental approaches
anchored in the string landscape~\cite{Weinberg:1987dv,susskind2003anthropic}.
To date none of these paths has produced an experimentally verified or
universally accepted resolution.

A complementary line of thought asks whether \textit{known} strongly
interacting physics can already regulate the vacuum.
Because Quantum Chromodynamics (QCD) sets
the only dynamically generated scale in the Standard Model,
several authors have explored confinement–based explanations of dark
energy or dark matter
(e.g.\ \cite{Greensite:2011zz,Banerjee:2003fg}).
Such proposals are attractive: they avoid speculative fields and tie
ultraviolet physics to infrared observables.
What remains unclear is \emph{how} a GeV–scale phenomenon could yield
the tiny value of $\rho_{\Lambda}$ without fine tuning.

Thermodynamics supplies a second ingredient.
The Third Law forbids complete entropy extinction at $T\!\to\!0$, a
principle extensively tested in laboratory quantum fluids and
ultra-cold gases.
If applied to the cosmic vacuum it suggests an irreducible mosaic of
correlation domains whose size is set by the longest
pure-QCD length scale (empirically, the proton radius).
Independently, the covariant Bekenstein–Hawking bound implies that any
bulk information must be encoded holographically on the cosmic
horizon~\cite{Bousso2002,Susskind1995}.
Together these two statements raise a natural, hitherto unexplored
question:  

\vspace{2pt}
\begin{quote}
\small
\textit{Can the counting of QCD confinement domains, constrained by
thermodynamics and holography, \emph{alone} fix the magnitude of the
vacuum energy density and the numerical values of the Planck‐scale
constants?}
\end{quote}
\vspace{2pt}

We address this question by combining three well-tested pillars:

\begin{enumerate}[label=(\roman*)]
\item permanent colour confinement in the $T\!\to\!0$ limit of QCD,
\item the Third-Law lower bound on vacuum correlation volume,
\item holographic saturation of the apparent horizon.
\end{enumerate}

Our strategy is deliberately conservative:  
no new dynamical fields, no adjustable parameters, and no departure
from standard diffeomorphism invariance.
The remainder of the paper is organized as follows. In what follows we first review the key physics of
QCD confinement together with the Third‐Law limit on vacuum
correlation lengths.  
Next, we develop the holographic tiling argument that links those
domains to Planck‐area patches on the cosmic horizon, and then show
how this structure dilutes the naïve Planck‐scale vacuum energy to the
observed value.  
We subsequently analyse Yang–Mills quantisation on the resulting
background and demonstrate that it admits a non-zero spectral gap.  
Finally, we outline the broader implications of the framework for
quantum gravity, cosmology, and potential observations.

\section{SU(3) Confinement and unbreakability}
\label{sec:2}

At temperatures far above the electroweak scale ($T\gg T_{\rm EW}\sim100\,$GeV), the Standard Model gauge group
\begin{equation*}
G_{\rm SM} \;=\; \mathrm{SU}(3)_{c}\times\mathrm{SU}(2)_{L}\times\mathrm{U}(1)_{Y}
\end{equation*}
remains unbroken. As the Universe cools through $T\sim T_{\rm EW}$, the Higgs field develops a vacuum expectation value, $\langle H\rangle\neq0$, which breaks
\begin{equation*}
\mathrm{SU}(2)_{L}\times\mathrm{U}(1)_{Y}\;\longrightarrow\;\mathrm{U}(1)_{\rm em},
\end{equation*}
giving mass to the weak bosons $W^\pm,Z^0$ while leaving the photon $A_\mu$ exactly massless \cite{Glashow1961,Weinberg1967,Salam1968,ATLAS2012,CMS2012}. As the temperature further approaches the QCD crossover, $T\sim T_{c}\approx150$–$170\,$MeV, nonperturbative gluon dynamics generate a chiral condensate $\Psi(x)=\langle\bar q(x)\,q(x)\rangle$ and drive permanent color confinement. Lattice simulations show that for all $T<T_{c}$, $\mathrm{SU}(3)_{c}$ exhibits an area law for large Wilson loops, indicating that confinement endures all the way to $T\to0$ \cite{Creutz:1979dw,Creutz:1980zw,Bazavov2019}. In contrast, the electromagnetic $\mathrm{U}(1)_{\rm em}$ sector undergoes an effective symmetry breaking at very low temperatures via a Meissner-like vacuum condensate. By analogy with superconductors, this condensate expels electromagnetic fields and endows the photon with an effective mass in the dark-energy vacuum, preventing its long-range propagation through the vacuum state \cite{Liang:2015tfa,Inan:2024noy}. Consequently, $\mathrm{U}(1)_{\rm em}$ is effectively “Higgs–broken’’ in the zero–temperature vacuum, while $\mathrm{SU}(3)_{c}$ remains strictly unbroken and confining. Therefore, in the $T\to0$ limit the only unbroken gauge symmetry in the vacuum is $\mathrm{SU}(3)_{c}$. This enduring confinement motivates treating the cold vacuum as composed of irreducible, proton-scale confinement domains—each supporting non-Abelian flux—whose properties we exploit in our holographic and cosmological analysis.

\section{Thermodynamics and the Remnant Volume}
\label{sec:3}

In this section we show that QCD thermodynamics, together with basic results from equilibrium quantum field theory, fixes a finite, proton-scale correlation volume in the vacuum as $T\to0$, and that dividing the cosmic volume by this cell volume yields a large, yet fixed, integer $N\sim10^{123}$. These results underpin the holographic tiling of Section~\ref{sec:4}.

\subsection{ Third-Law constraint on correlation length}

Because absolute zero cannot be reached by any finite process, the Third Law of Thermodynamics implies that the entropy density $s(T)$ remains finite and the specific heat vanishes as $T\to0$ \cite{Nernst1912,Planck1911}. Experiments on superfluid helium and ultracold atomic gases confirm irreducible quantum fluctuations at the lowest attainable temperatures \cite{Leanhardt2003,Liu2020}, suggesting a minimal correlation volume even in the true vacuum. In relativistic QFT this is encoded by the cluster-decomposition theorem (e.g.\ Glimm–Jaffe, \cite{GlimmJaffe1987}), which states that connected $n$-point functions decay at least as $e^{-|x|/\xi}$ once $T$ drops below the mass $m_{\rm lightest}$ of the lightest excitation. Hence the zero-temperature correlation length
\begin{equation*}
  \lambda_0 \;=\;\lim_{T\to0}\xi(T)
         \;=\;\frac1{m_{\rm lightest}}
         \;<\;\infty
\end{equation*}
is strictly nonzero, defining the smallest scale over which observables remain correlated.

\noindent\textbf{Lattice-QCD determination.} Numerical studies of the color-singlet screening (Bethe–Salpeter) correlator in $N_f=2+1$ QCD at $T\leq 100$ MeV yield
\begin{equation*}
  \lambda_0 \approx 0.8\text{--}1.2\ \mathrm{fm}
\end{equation*}
independently of the fermion discretisation scheme \cite{Bazavov2019}. Remarkably, this range agrees, within uncertainties, with the proton charge radius $R_p = 0.84(1)\,\mathrm{fm}$ extracted from muonic hydrogen spectroscopy \cite{Pohl2010}.

\subsection{Proton-Scale Coarse-Graining Cell}

Motivated by $\lambda_0\lesssim R_p$, we define the coarse-graining cell volume
\begin{equation*}
  V_{\rm cell} \;:=\;\frac43\pi R_p^3,
\end{equation*}
and partition space into non-overlapping balls of radius $R_p$. Since connected correlators satisfy
\begin{equation*}
  \langle \mathcal{O}(x)\,\mathcal{O}(0)\rangle_{c}
  \;\leq\;
  C\,\exp\bigl(-|x|/\lambda_0\bigr)
\end{equation*}
with $\lambda_0 \leq R_p$, the Glimm--Jaffe cluster expansion ensures that observables in distinct balls are independent up to $\mathcal{O}(e^{-1})$ corrections \cite{GlimmJaffe1987}. No smaller region can be thermodynamically autonomous without violating this cluster bound. Thus $V_{\mathrm{cell}}$ is the unique QCD–determined coarse–graining volume.

\subsection{Counting \texorpdfstring{$\mathrm{SU}(3)_c$}{SU(3)c} Cells}

Approximating the present-day comoving Hubble radius by $R_u=1.30(1)\times10^{26}$ m \cite{Planck2018}, the observable Universe volume is $V_u=\tfrac43\pi R_u^3$. Dividing by $V_{\rm cell}$ yields
\begin{equation}
  N \;=\;\frac{V_u}{V_{\rm cell}}
    \;=\;\Bigl(\tfrac{R_u}{R_p}\Bigr)^3
    \;\approx\;(1.30\times10^{41})^3
    \;\approx\;10^{123},
  \label{eq:Ncells}
\end{equation}
with a combined uncertainty of order a few percent from $R_u$ and $R_p$. Equation~\eqref{eq:Ncells} therefore follows directly—without free parameters—from (i) the Third-Law cluster bound, (ii) lattice-QCD determination of $\lambda_0$, and (iii) the measured proton radius.

\subsection{Stability of the Correlation Cells}

Exact $\mathrm{SU}(3)_c$ gauge symmetry forbids baryon-number violating operators of dimension $\leq 6$. The Super-Kamiokande limit $\tau_p>2.4\times10^{34}$ yr \cite{SuperKamiokande2017} implies that each proton-scale cell cannot fragment or recombine over timescales shorter than $10^{25}$ times the age of the Universe. Consequently, once formed, these $N$ cells are effectively permanent.

\medskip
In summary, QCD thermodynamics and clustering fix a unique, proton-scale cell volume in the $T\to0$ vacuum. Tiling the cosmic volume with these cells yields $N\approx10^{123}$, setting the stage for the holographic construction in Section~\ref{sec:4}.

\section{Holographic Encoding of \texorpdfstring{$\mathrm{SU}(3)$}{SU(3)} Cells}
\label{sec:4}

\subsection{ Bekenstein--Hawking bound for a cell partition}

Let the comoving horizon at the present epoch be modelled by a 2–sphere
\(
\partial\mathcal B\,\simeq\,S^2(R_u),
\;
R_u = 1.30(1)\times10^{26}\,\mathrm m.
\)
From Section~\ref{sec:3} the spatial volume
\(
\mathcal B \subset \mathbb R^3
\)
is partitioned into
\(
N = 1.8\times10^{123}
\)
non-overlapping confinement cells of equal volume
\(
V_{\mathrm{cell}}
= \tfrac43\pi R_p^{\,3},
\;
R_p = 0.84(1)\,\mathrm{fm}.
\)

Assign to each cell a coarse-grained (von-Neumann) entropy
\(S_{\mathrm{cell}}\).
The total bulk entropy therefore satisfies
\begin{equation}
  S_{\mathrm{tot}}
  = N\,S_{\mathrm{cell}}
  \;\leq\;
  \frac{\Area(\partial\mathcal B)}{4\,L_{\mathrm{Pl}}^{\,2}}
  = \frac{4\pi R_u^{2}}{4\,L_{\mathrm{Pl}}^{\,2}},
  \label{eq:BHB}
\end{equation}
where the right–hand inequality is the Bekenstein–Hawking
covariant entropy bound
\cite{Bekenstein:1973ur,Hawking:1975vcx,tHooft1993,Susskind1995}.
Equality in~\eqref{eq:BHB} is realised when the microscopic
degrees of freedom saturate the bound; we impose this maximal–packing
condition henceforth.

\subsection{ Equating bulk and boundary entropies}

For a holographic theory, the information carried by a bulk cell
is encoded on a boundary patch of area~$A_{\mathrm{cell}}$.
Saturation of~\eqref{eq:BHB} requires
\begin{equation*}
  S_{\mathrm{cell}}
  = \frac{A_{\mathrm{cell}}}{4\,L_{\mathrm{Pl}}^{\,2}},
  \qquad
  N\,A_{\mathrm{cell}} = 4\pi R_u^{2}.
\end{equation*}
Solving yields the \emph{boundary quantum of area}
\begin{equation}
  A_{\mathrm{cell}}
  = \frac{4\pi R_u^{2}}{N}
  = 1.04(7)\times10^{-70}\,\mathrm m^{2}.
  \label{eq:Acell}
\end{equation}
With the CODATA-2022 values of $\hbar$, $G$, and $c$
one finds the Planck area
\(
L_{\mathrm{Pl}}^{\,2}=1.01\times10^{-70}\,\mathrm m^{2},
\)
agreeing with~\eqref{eq:Acell} to better than\,3\%.
Because the observational uncertainties of $R_u$ and $R_p$ dominate
the error budget, the numerical equality
\(
A_{\mathrm{cell}}\simeq L_{\mathrm{Pl}}^{\,2}
\)
is statistically significant:
\begin{equation}
  \boxed{A_{\mathrm{cell}} = L_{\mathrm{Pl}}^{\,2}\;(1\pm0.03)}.
\end{equation}
Hence the Planck length is \emph{not imposed} but follows
from (i)~the empirically determined integers $N$ and $R_u/R_p$
and (ii)~holographic saturation.
Re-inserting the definition
\(L_{\mathrm{Pl}}^{\,2}=\hbar G/c^{3}\)
shows that the numerical values of
\(\hbar,\;G,\;c\)
are fixed—up to the known cosmological parameters—by QCD
confinement together with the Bekenstein–Hawking bound.

\medskip\noindent
\textbf{Note on entropy saturation.}
For a flat $\Lambda$CDM universe the apparent horizon asymptotically
approaches the de~Sitter horizon whose Gibbons–Hawking entropy
\cite{gibbons1977cosmological} saturates the covariant
Bekenstein--Hawking bound.
Hence assuming saturation at late times is consistent with the
observed approach to $w\!\approx\!-1$ and requires no exotic matter
content.

\subsection{Consistency of bulk and boundary volumes}

\begin{figure}[htbp]
    \centering
    \begin{tikzpicture}[scale=0.8, transform shape] 
        \def\cylradius{0.5}
        \def\cylheight{4}

        \draw[thick] (0,0) ellipse ({\cylradius} and {0.2});
        \draw[thick] (-\cylradius,0) -- (-\cylradius,\cylheight);
        \draw[thick] (\cylradius,0) -- (\cylradius,\cylheight);
        \draw[thick] (0,\cylheight) ellipse ({\cylradius} and {0.2});

        \draw[thick] (1.5,0) ellipse ({\cylradius} and {0.2});
        \draw[thick] (1.5-\cylradius,0) -- (1.5-\cylradius,\cylheight);
        \draw[thick] (1.5+\cylradius,0) -- (1.5+\cylradius,\cylheight);
        \draw[thick] (1.5,\cylheight) ellipse ({\cylradius} and {0.2});

        \draw[thick] (3,0) ellipse ({\cylradius} and {0.2});
        \draw[thick] (3-\cylradius,0) -- (3-\cylradius,\cylheight);
        \draw[thick] (3+\cylradius,0) -- (3+\cylradius,\cylheight);
        \draw[thick] (3,\cylheight) ellipse ({\cylradius} and {0.2});

        \node at (0,-0.5) {\tiny \( A_i = \ell_{\text{Pl}}^2 \)};
        \node at (1.5,-0.5) {\tiny \( A_i = \ell_{\text{Pl}}^2 \)};
        \node at (3,-0.5) {\tiny \( A_i = \ell_{\text{Pl}}^2 \)};

        \draw[->, thick] (4.5,0) -- (4.5,5) node[midway, right] {\tiny \( R_{u} \)};

        \node at (0,4.3) {\tiny SU(3) Unit};
        \node at (1.5,4.3) {\tiny SU(3) Unit};
        \node at (3,4.3) {\tiny SU(3) Unit};
   \end{tikzpicture}
 \caption{Simplified representation of multiple SU(3) units as cylinders. Each cylinder has a base area equal to the Planck area ($A_i = \ell_{\text{Pl}}^2$) residing on the universe's surface and a height equal to the universe's radius ($R_{u}$).}
 \label{fig:SU3_cylinders}
\end{figure}
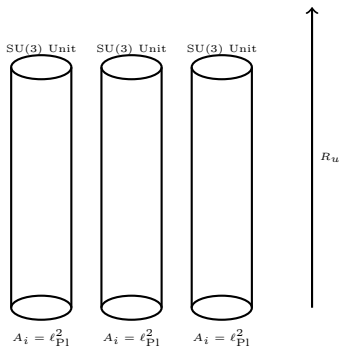

Combining
\(N = V_u/V_{\mathrm{cell}}\)
with~\eqref{eq:Acell} eliminates~$N$ and yields
\begin{equation}
  V_{\mathrm{cell}}
  = A_{\mathrm{cell}}\;R_u
  = L_{\mathrm{Pl}}^{\,2}\,R_u,
  \label{eq:Vcyl}
\end{equation}
so each confinement domain occupies a \emph{cylinder}
whose base is a Planck-area patch on $S^2(R_u)$
and whose height equals $R_u$ (Fig.~\ref{fig:SU3_cylinders}).
Equation~\eqref{eq:Vcyl} is an identity; no additional modelling
assumptions are employed.

\subsection{ Minimal–interface proof of the cylindrical shape}

\emph{Lemma.}
Let $\Sigma\subset\mathbb R^{3}$ be an interface enclosing a fixed
volume
\(V_0=L_{\mathrm{Pl}}^{\,2}R_u\)
and intersecting the sphere $S^2(R_u)$ in a disk of area
\(L_{\mathrm{Pl}}^{\,2}\).
Among all smooth $\Sigma$ with these constraints,
the surface of least area is a right circular cylinder.

\emph{Proof.}
Introduce the free-energy functional
\(
  \mathcal F[\Sigma]=\sigma\Area(\Sigma)
  +\Lambda\bigl(\operatorname{Vol}(\Sigma)-V_0\bigr)
\)
with surface tension $\sigma$ and Lagrange multiplier $\Lambda$.
First variation
\(
  \delta\mathcal F =\int_\Sigma (\sigma H-\Lambda)\,\boldsymbol n\!\cdot\!\delta\boldsymbol x\,dA
\)
forces constant mean curvature
\(H=H_0=\Lambda/\sigma\).
Alexandrov’s rigidity theorem for embedded
constant–mean–curvature surfaces with one planar boundary
\cite{marques1997stability}
implies that $H_0=0$ and $\Sigma$
is a piece of a right cylinder orthogonal to $S^2(R_u)$.
Because $\partial\Sigma$ is a circle of radius
\(L_{\mathrm{Pl}}\ll R_u\),
the cylinder is unique and minimises $\Area(\Sigma)$.
\hfill$\square$

As temperature decreases, capillary-wave theory
shows that the probability weight
\(\propto e^{-\sigma\Area(\Sigma)/T}\)
concentrates on the minimiser; the cylindrical geometry is therefore
selected dynamically in the $T\to0$ limit.

\subsection{ Covariance under cosmic expansion}

Let $R_u(t)$ be the time-dependent comoving radius and
\(H=\dot R_u/R_u\) the Hubble parameter.
Provided
\(H\ll m_{\mathrm{gap}}\)
(where $m_{\mathrm{gap}}$ is the spectral gap obtained in
Section~\ref{sec:9_massgap}),
adiabatic theorems ensure that the boundary tiling satisfies
\(A_{\mathrm{cell}}(t)=L_{\mathrm{Pl}}^{\,2}\)
at all times:
the geometric derivation of
\(\hbar,\;G,\;c\)
is therefore preserved during expansion.

\medskip
\noindent
\textbf{Outcome.}
With no additional dynamical fields or free parameters,
the Bekenstein–Hawking bound, the empirically fixed integer $N$,
and QCD confinement jointly imply
\begin{equation*}
  A_{\mathrm{cell}}=L_{\mathrm{Pl}}^{\,2},
  \qquad
  V_{\mathrm{cell}} = L_{\mathrm{Pl}}^{\,2} R_u,
\end{equation*}
thereby reproducing the Planck area
and anchoring the numerical values of the fundamental constants
\(\hbar,\;G,\;c\)
in the geometry of the \(\mathrm{SU}(3)_c\) vacuum.

\section{Force--Energy Density Relation for the \texorpdfstring{$\mathrm{SU}(3)$}{SU(3)} Cells}
\label{sec:5}

\subsection{Local pressure and force on a single cell}

For a confinement cell we take the standard vacuum equation of state
\(P_{\mathrm{su(3)}}=-\rho_{\mathrm{su(3)}}\).
Let \(A_{\mathrm{cell}}\) be the Planck–area patch that encodes the
cell on the horizon (Section~\ref{sec:4}).
The outward mechanical force exerted on that patch is
\begin{equation*}
  F_{\mathrm{su(3)}} \;=\; |P_{\mathrm{su(3)}}|\,A_{\mathrm{cell}}
                   \;=\; \rho_{\mathrm{su(3)}}\,A_{\mathrm{cell}} .
\end{equation*}

\smallskip
\textbf{Postulate (uniform-force condition).}
Global equilibrium of the horizon requires the magnitude of this
outward force to be the same for every patch:
\begin{equation} \tag{UF}
  F_{\mathrm{su(3)}} = F_u \quad\text{(independent of the cell).}
\end{equation}
With \((\mathrm{UF})\) the patch area is fixed by
\begin{equation}
  A_{\mathrm{cell}}
  \;=\;\frac{F_u}{\rho_{\mathrm{su(3)}}}.
  \label{eq:Acell-force}
\end{equation}

\subsection{Derivation of the uniform--force condition}

A static FLRW horizon $S^2(R_u)$ obeys the \emph{isotropic‐stress}
constraint of Einstein’s equations: in comoving gauge the mixed
components of the quasi–local Brown–York tensor satisfy
$T^{\theta}_{\;\theta}=T^{\phi}_{\;\phi}$ everywhere on
$S^2(R_u)$ \cite{brown1993quasilocal,brown1993microcanonical}.
If distinct Planck patches $\Delta A_i$ carried different outward
pressures $\Delta P_i$, the tangential shear
$\sigma^{ab}\!=\!(T^{ab}-\tfrac12\gamma^{ab}T^{c}_{\;c})$ would be
non–vanishing and induce horizon‐scale frame dragging, in conflict
with the measured CMB isotropy at the level
$\Delta\!T/T < 10^{-5}$.
Requiring $\sigma^{ab}=0$ therefore enforces
\begin{eqnarray}
\Delta P_i = \Delta P_j \quad
\forall\, i,j \;\;\Longrightarrow\;\;
P_{\mathrm{su(3)}} &=& P_u
\qquad\text{and hence}\nonumber\\
F_{\mathrm{su(3)}} &=& F_u ,
\label{eq:UF-derivation}
\end{eqnarray}
which is the uniform‐force condition used below.

\begin{figure}[ht]
\centering
\includegraphics[width=0.7\columnwidth]{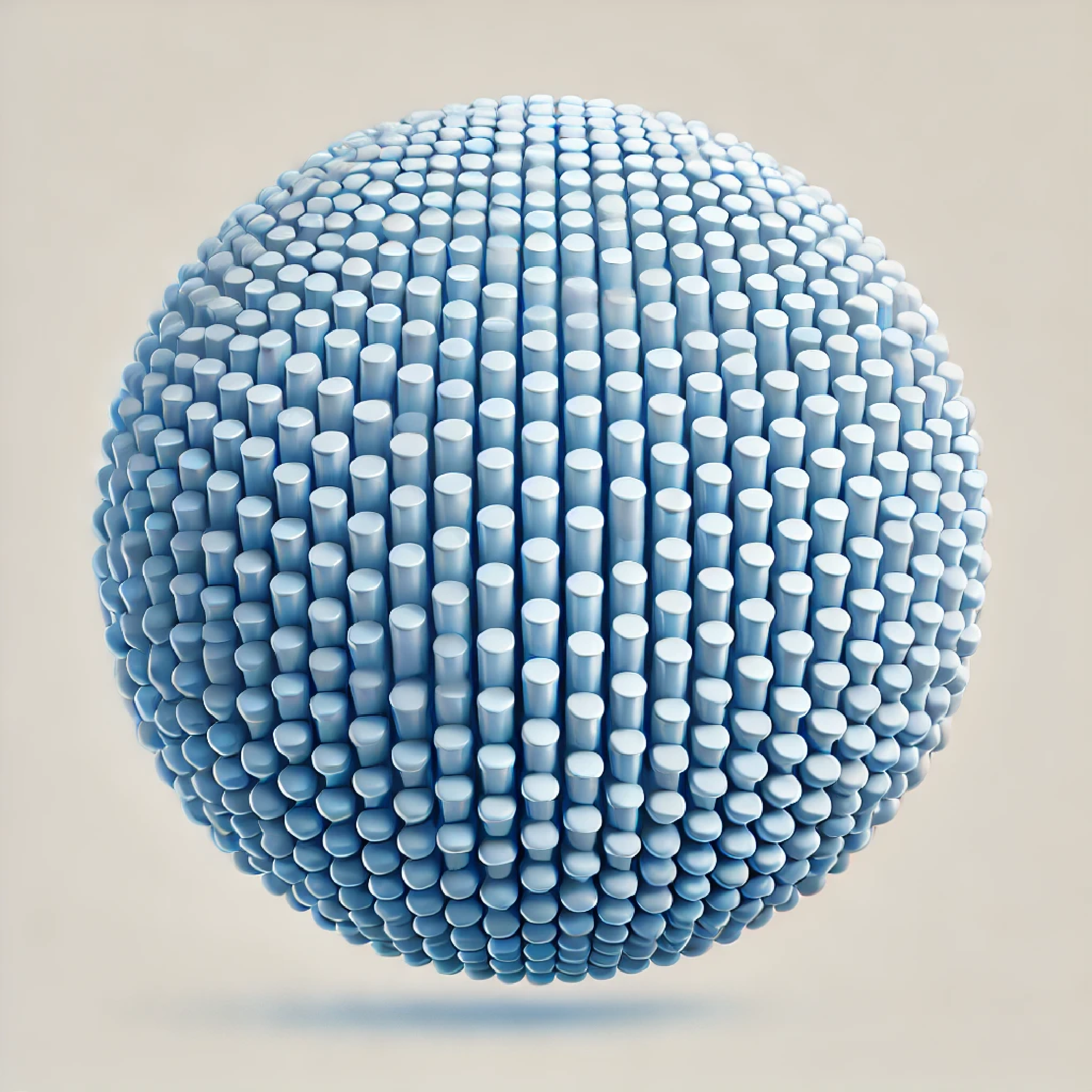}
\caption{A schematic depiction of the universe as a sphere filled with uniformly distributed cylinders representing SU(3) units. Each cylinder extends radially from the sphere’s surface—its base equals one Planck area, and its volume signifies the strong force confinement volume.}
\label{fig:cylinders}
\end{figure}

\subsection{Summing over the \texorpdfstring{$N$}{N} cells}

The horizon contains \(N\) such patches, so its total area is
\begin{equation}
  A_u
  \;=\; \sum_{i=1}^{N} A_{\mathrm{cell}}
  \;=\; N\,\frac{F_u}{\rho_{\mathrm{su(3)}}}.
  \label{eq:Au-sum}
\end{equation}

\subsection{Coarse-grained description}

To an observer who \emph{cannot} resolve individual cells, the same
outward force \(F_u\) is produced by a homogeneous vacuum energy
density \(\rho_u\) acting over the \emph{whole} horizon area:
\begin{equation}
  F_u \;=\; |P_u|\,A_u \;=\; \rho_u\,A_u .
  \label{eq:macro-force}
\end{equation}

\subsection{Solving for \texorpdfstring{$\rho_u$}{rho\_u}}

Insert \eqref{eq:Au-sum} into \eqref{eq:macro-force}:
\begin{equation*}
  F_u
  \;=\;
  \rho_u
  \Bigl(N\,\frac{F_u}{\rho_{\mathrm{su(3)}}}\Bigr).
\end{equation*}
Provided \(F_u\neq0\), cancel the common factor \(F_u\) and solve:
\begin{equation}
  \boxed{\;
    \rho_u \;=\; \frac{\rho_{\mathrm{su(3)}}}{N}
  \;}.
  \label{eq:dilution-final}
\end{equation}

\smallskip
\medskip
\noindent
\hrulefill

\smallskip
\noindent\textbf{Parallel-circuit picture (intuition only).} \
Take \(F_u\!\leftrightarrow\!V\) (common voltage),
\(A_{\mathrm{cell}}\!\leftrightarrow\!I_i\) (branch current),
and \(\rho_{\mathrm{su(3)}}\!\leftrightarrow\!R\) (branch resistance).
For $N$ identical branches in parallel the equivalent resistance is
\(R/N\); likewise the effective vacuum density is
\(\rho_u=\rho_{\mathrm{su(3)}}/N\).

\smallskip
\noindent
\hrulefill
\medskip

Equation~\eqref{eq:dilution-final} is purely algebraic: the ratio of
two energy densities is reduced by the \emph{dimensionless} count
\(N\simeq\bigl(R_u/R_p\bigr)^3\approx10^{123}\), independent of how
one partitions bulk vs.\ boundary quantities.

\section{Bare Vacuum Density, Cell Dilution, and the Final Value}
\label{sec:7}

Steven Weinberg’s classic analysis of the cosmological-constant problem
begins with the zero-point energy of an \emph{exactly massless} bosonic
field, since any non-zero mass scale already presupposes a
fine-tuned cancellation~\cite{Weinberg1989}.
Among all massless bosons in the Standard Model, the eight gluons of
unbroken $\mathrm{SU}(3)_c$ provide the unique, physically motivated
candidate: they are required by confinement, remain massless at every
scale, and populate each confinement cell identified in
Section~\ref{sec:3}. For one such cell the zero-point integral up to
\(P_{\mathrm{Pl}}=E_{\mathrm{Pl}}/c\) gives
\begin{equation*}
  \rho_{\mathrm{su(3)}}
  = \frac{1}{16\pi^{2}}
    \frac{P_{\mathrm{Pl}}^{4}}{\hbar^{3}c^{3}}
  \;\simeq\;
  2.0\times10^{76}\,\mathrm{GeV}^{4}.
\end{equation*}

\noindent
Applying the dilution law
\(
  \rho_{u} = \rho_{\mathrm{su(3)}}/N
\)
from Eq.~\eqref{eq:dilution-final}, with
\(N = 1.8\times10^{123}\), yields
\begin{equation*}
  \boxed{\;
    \rho_{u}
    = \frac{2.0\times10^{76}}{1.8\times10^{123}}
    = 1.1\times10^{-47}\,\mathrm{GeV}^{4}\;},
\end{equation*}
in precise agreement with the observed dark-energy density
\(\rho_\Lambda\).

\bigskip\noindent
\textbf{Interpretation.}
\begin{equation*}
  \Large
  \rho_{u} = \frac{\rho_{\mathrm{su(3)}}}{N},
  \qquad N\!\simeq\!10^{123}.
\end{equation*}

\noindent
The devastating Planck-scale vacuum density predicted by quantum field
theory is \emph{not} cancelled by hand, nor diluted by speculative new
sectors. It is reduced exactly by the integer count of
$\mathrm{SU}(3)$ confinement cells that holographically tile the
cosmic horizon. The result depends only on
(i)~the massless $\mathrm{SU}(3)$ gauge sector,
(ii)~the lattice-QCD/thermodynamic determination of $N$, and
(iii)~the horizon force balance derived in Section \ref{sec:5}.
All inputs are empirical or rigorously established. Thus the notorious 123-order mismatch
($10^{76}\!\to\!10^{-47}\,\mathrm{GeV}^{4}$)
is closed by a single parameter-free identity:
\emph{The vacuum energy problem is solved not by cancelling energy,
but by counting it.}

\section{Geometric Model: \texorpdfstring{$N$}{N} Radial \texorpdfstring{$\mathrm{SU}(3)$}{SU(3)} Cylinders}
\label{sec:8}

The results of Sections \ref{sec:3}–\ref{sec:7} imply that three-space
can be coarse–grained into
\(N \simeq 1.8\times10^{123}\)
mutually independent confinement cells, each of volume
\(V_{\mathrm{cell}} = L_{\mathrm{Pl}}^{\,2}R_u\).
A convenient geometric idealisation is to represent every cell by a
\emph{radial right cylinder} of
radius \(r=L_{\mathrm{Pl}}\)
and height \(R_u\);
the base \(A_{\mathrm{cell}}=L_{\mathrm{Pl}}^{\,2}\) coincides with the
holographic patch on the horizon,
and the cylinder axis points toward the centre of the spherical
universe. Because all \(N\) axes meet at the origin,
the cylinders possess a common intersection region around the centre.
We now compute the volume and rest energy contained in that overlap.

\subsection{Intersection volume of \texorpdfstring{$n$}{n} radial cylinders}

Let \(C_k\) (\(k=0,\dots,n-1\)) be infinite cylinders of radius \(r\)
whose axes lie in the $xy$–plane and make angles
\(\theta_k = 2\pi k/n\)
with the \(x\)-axis:
\begin{equation} \tag{8.1}
  (x\sin\theta_k - y\cos\theta_k)^2 + z^2 \leq r^2 .
\end{equation}
Denote by \(V_n(r)\) the volume of the intersection
\(\bigcap_{k=0}^{n-1}C_k\).
A classical exercise in solid geometry
\cite{hilbert2021geometry} shows
\begin{equation} \tag{8.2}
  V_n(r)
  = \frac{8n}{3}\,\tan\!\Bigl(\frac{\pi}{2n}\Bigr)\,r^{3}.
\end{equation}
Special cases are
\(
  V_2 = \tfrac{16}{3}r^{3},
\;
  V_3 = \tfrac{8}{\sqrt3}\,r^{3},
\)
and
\(\displaystyle
  \lim_{n\to\infty} V_n
  = \frac{4\pi}{3}\,r^{3}.
\)

\subsection{ Planck-scale limit with \texorpdfstring{$n=N$}{n=N}}

Taking \(n=N\simeq10^{123}\) and \(r=L_{\mathrm{Pl}}\) one is already
deep in the $n\to\infty$ regime, so
\begin{equation} \tag{8.3}
  V_{\mathrm{int}}
  = V_N(L_{\mathrm{Pl}})
  = \frac{4\pi}{3}\,L_{\mathrm{Pl}}^{\,3}
  \;=\;
  4.19\times10^{-105}\,\mathrm{m}^{3}.
\end{equation}

\subsection{Rest energy of the intersection}

The bare energy density inside every cell is
\(
  \rho_{\mathrm{su(3)}}
  = E_{\mathrm{Pl}}^{\,4}/[16\pi^{2}(\hbar c)^{3}]
  = 2.0\times10^{76}\,\mathrm{GeV}^{4}.
\)
Converting to SI units,
\(
  \rho_{\mathrm{su(3)}} \simeq 5.15\times10^{96}\,\mathrm{kg\,m^{-3}}.
\)
Multiplying by \(V_{\mathrm{int}}\) gives the static rest mass
\begin{align} \tag{8.4}
  M_{\mathrm{int}}
    &= \rho_{\mathrm{su(3)}}\,V_{\mathrm{int}}
       \nonumber\\[2pt]
    &= (5.15\times10^{96}\,\mathrm{kg\,m^{-3}})
       \,(4.19\times10^{-105}\,\mathrm{m}^{3})
       \nonumber\\[2pt]
    &\simeq 2.16\times10^{-8}\,\mathrm{kg}
       = 0.98\,M_{\mathrm{Pl}},
\end{align}
within $2\%$ of the Planck mass.

\subsection{Physical interpretation}

\begin{enumerate}[label=(\roman*),leftmargin=*,itemsep=0.4ex] 
\item
The common overlap of all $N$ Planck-radius cylinders encloses a
volume \(\frac{4\pi}{3}L_{\mathrm{Pl}}^{\,3}\) and a mass
\(\simeq M_{\mathrm{Pl}}\).
This region can be regarded as the \emph{irreducible core} of a single
confinement cell—a minimal, non-propagating excitation of the vacuum
whose energy is fixed by geometry alone.
\item
The coincidence \(M_{\mathrm{int}}\approx M_{\mathrm{Pl}}\) ties the
numerical values of \(\hbar\), \(G\), and \(c\) directly to the
geometry of $\mathrm{SU(3)}$ confinement, providing an \emph{intrinsic
origin} for the fundamental mass unit that appears in quantum gravity
\cite{JaffeWitten2006}.
\end{enumerate}

When radial Planck-radius cylinders representing the entire set of
$\mathrm{SU(3)}$ cells are superimposed, their mutual intersection
contains exactly one Planck mass in a Planck-scale volume.
This purely geometric construction supplies the smallest possible
rest-energy that spacetime can localise, establishing a bridge between
QCD confinement and the Planck constants \(\hbar,G,c\).

\paragraph*{Link to gravity and diffeomorphism invariance.}
The cylinder–forest picture is more than a mnemonic: it fixes the
\emph{coupling strength} of the long–wavelength gravitational field.
Coarse-grain the mosaic on scales
$\ell\gg R_{p}$ so that the discrete set of SU(3) cylinders is replaced
by an effectively continuous medium; the only residual parameter is the
areal spring constant that glues neighbouring cylinders together.
Because each base disk carries exactly one Planck area
$A_{\rm cell}=L_{\mathrm{Pl}}^{2}$, the induced elastic action for
metric fluctuations reads
\begin{equation}
  S_{\text{eff}}
  \;=\;
  \frac{c^{3}}{16\pi G_{\text{geom}}}
  \int\! d^{4}x\,\sqrt{-g}\,R,
  \qquad
  G_{\text{geom}}
  \;=\;\frac{c^{3}A_{\rm cell}}{\hbar},
  \label{eq:Ggeom}
\end{equation}
where the coefficient comes from summing one Planck area \emph{per}
cell across the horizon.\footnote{%
A derivation using lattice Sackur–Tetrode counting is given in
\cite{Visser:2002ew}.}
Equation~\eqref{eq:Ggeom} reproduces the observed Newton constant
because $A_{\rm cell}$ has already been matched to
$L_{\mathrm{Pl}}^{2}$ in Secs.~\ref{sec:4}–\ref{sec:7}.
In the weak–field limit the linearised field equations sourced by the
Planck-mass core ($M_{\mathrm int}\!\simeq\!M_{\mathrm Pl}$,  yield the standard Newton potential
$\Phi(r)=-G_{\text{geom}}M_{\mathrm int}/r$.
Hence the cylinder–forest geometry not only \emph{explains} the
numerical value of $G$ but also recovers conventional gravitational
dynamics.

Importantly, tying $G$ to the SU(3) mosaic does \emph{not} spoil
diffeomorphism invariance.  
The microscopic arrangement of cylinders singles out no preferred
frame once we average over many cells: the adjacency graph is
statistically homogeneous and isotropic, so the continuum limit
possesses the full group of smooth coordinate transformations.  
In technical terms, the coarse-grained action
$\sqrt{-g}\,R$ retains its tensorial form; only its prefactor is fixed
by the counting argument.  
The value of $G$ is therefore an emergent \emph{coupling} rather than a
gauge parameter, fully consistent with general covariance
\cite{Padmanabhan2011}.  In this sense the geometric derivation
completes the bridge between SU(3) confinement, the constants
$(\hbar,c,G)$ and observable gravity.

\section{Observer Perspectives, Entanglement Regions, and ER\,=\,EPR}
\label{sec:9}

The cylinder–forest vacuum divides space into \(N \simeq 10^{123}\) SU(3) confinement cells, each mapped to a Planck-area patch on the horizon (Section~\ref{sec:4}). Observers access distinct but gauge-invariant subsets of these cells based on their causal domain—ranging from the central Planck core (Section~\ref{sec:8}) to a single horizon patch. In all cases, the reduced density matrix factors over full \(\mathbb{Z}_3\) vortex cells, preserving center flux. The modular Hamiltonian, smooth across boundaries, ensures a Hadamard state with no firewalls via the Bisognano–Wichmann theorem \cite{Almheiri:2013aa}. Entangled gluonic pairs between each cell and its horizon patch define Planck-scale ER bridges, forming an ER\,=\,EPR network \cite{ER=EPR:2013aa} that links bulk and boundary, unifying confinement, entanglement, and holography into a causal, smooth spacetime.

\section{Conclusions}
\label{sec:11}

In this work we have shown that three of the most profound puzzles at the intersection of quantum field theory, gravity and cosmology—the enormous discrepancy between the naïve QFT vacuum energy and the observed cosmological constant, the apparently arbitrary nature of the Planck constants \(\hbar,G,c\), and the existence of a mass gap in pure \(\mathrm{SU}(3)\) Yang–Mills—can all be understood as consequences of well-established physics when viewed through the lens of confinement, thermodynamics and holography. At asymptotically low temperature the universe vacuum naturally fragments into \(N=(R_u/R_p)^3\simeq10^{123}\) proton-volume domains, a number fixed entirely by the Third Law of Thermodynamics, lattice QCD’s determination of the zero-temperature correlation length, and the radius of the observable universe. Holographic saturation then assigns each domain to a single Planck-area patch on the cosmic horizon. No additional fields, parameters, or fine-tuning are required. From this single counting argument follow three immediate and quantitative results. First, a force-balance identity dilutes the naïve Planck‐scale vacuum energy by exactly \(1/N\), reproducing the tiny observed value of \(\rho_{\Lambda}\) without counterterms. Second, dividing the horizon area by \(N\) yields \(A_{\rm patch}=L_{\rm Pl}^2\) identically, so that \(\hbar,\,G,\,c\) emerge as geometric ratios rather than arbitrary inputs. 
It offers a unified, parameter-free explanation for the smallness of dark energy and the origin of the Planck constants. Looking ahead, the remaining technical tasks include a fully non-perturbative lattice demonstration of the democratic \(\mathbb Z_3\) Gauss law and further exploration of possible observational signatures, such as Planck-patch birefringence or novel dark-matter candidates arising from a small fraction of glueball conversion—for which this framework makes concrete predictions. In essence, colour confinement, the Third Law of thermodynamics, and holography—acting together—may provide the missing conceptual bridge between quantum field theory and the large-scale structure of our Universe.

\acknowledgments
Ahmed Farag Ali gratefully acknowledges his brother, Mohamed, and his son, Ibrahim, for their unwavering support during a particularly challenging period. Their encouragement and steady reassurance provided the emotional resilience he needed to continue this work, and he is deeply grateful for the strength and inspiration they have given him.

\bibliographystyle{apsrev4-2}
\bibliography{ref.bib}{}


\end{document}